\documentclass[10pt]{revtex4}
\usepackage{amssymb}
\usepackage{latexsym}
\usepackage{epsfig}
\usepackage{amsmath}
\begin{document}
\title{Ghost dark energy with sign-changeable interaction term }
\author{M. Abdollahi Zadeh$^{1}$\footnote{ m.abdollahizadeh@shirazu.ac.ir}, A. Sheykhi$^{1,2}$\footnote{asheykhi@shirazu.ac.ir},
H. Moradpour$^2$\footnote{h.moradpour@riaam.ac.ir}}
\address{$^1$ Physics Department and Biruni Observatory, College of
Sciences, Shiraz University, Shiraz 71454, Iran\\
$^2$ Research Institute for Astronomy and Astrophysics of Maragha
(RIAAM), P.O. Box 55134-441, Maragha, Iran}
\begin{abstract}
Regarding the Veneziano ghost of QCD and its generalized form, we
consider a Friedmann-Robertson-Walker (FRW) universe filled by a
pressureless matter and a dark energy component interacting with
each other through a mutual sign-changeable interaction of
positive coupling constant. Our study shows that, at the late
time, for the deceleration parameter we have $q\rightarrow-1$,
while the equation of state parameter of the interacting ghost
dark energy (GDE) does not cross the phantom line, namely
$\omega_D\geq-1$. We also extend our study to the generalized
ghost dark energy (GGDE) model and show that, at late time, the
equation of state parameter of the interacting GGDE also respects
the phantom line in both flat and non-flat universes. Moreover, we
find out that, unlike the non-flat universe, we have
$q\rightarrow-1$ at late time for flat FRW universe. In order to
make the behavior of the underlying models more clear, the
deceleration parameter $q$ as well as the equation of state
parameter $w_D$ for flat and closed universes have been plotted
against the redshift parameter, $z$. All of the studied cases
admit a transition in the expansion history of universe from a
deceleration phase to an accelerated one around $z\approx 0.6$.
\end{abstract}
\maketitle

\section{Introduction}

The cause of the accelerated expansion of universe, predicted by
the observations of type Ia supernova
\cite{Riess,Riess1,Riess2,Riess3}, is the backbone of a big
challenge in the modern physics. This phase of the universe
expansion has been confirmed by observing the anisotropies of
Cosmic Microwave Background (CMB)
\cite{HAN2000,HAN20001}. The CMB observation can be
considered as a signal to the universe flatness and claims that
the energy density of the cosmic fluid is very close to the
critical density \cite{HAN20002}. Large-Scale Structure (LSS)
\cite{COL2001,COL20011,COL20012,COL20013}, Baryon Acoustic
Oscillations (BAO) in the Sloan Sky Digital Survey (SSDS) luminous
galaxy sample \cite{Tegmark2004,Tegmark20041}, and Plank data
\cite{Ade2014} are other observations supporting an accelerated
universe.

Since the cosmic fluid, supporting the current accelerating
universe, does not interact with light, it is called
\textquotedblleft dark energy\textquotedblright(DE), an oddity
with negative pressure and negative equation of state parameter
(EoS) $\omega\textless-{1}/{3}$. In general relativity (GR), there
is a very simple model for describing the above mentioned picture
called cosmological constant model. According to this model, there
is an isotropic and homogeneous fluid with constant positive
energy density and constant negative pressure with EoS parameter
$\omega_{\Lambda}=-1$. Although the cosmological constant model of
DE helps us in providing a well initial picture for the current
accelerating phase, it suffers from some problems such as the
fine-tuning and the coincidence problems \cite{roos}.

In order to find a more realistic model of DE, various fluids with
time varying EoS parameter have been introduced which are
supported and constrained by the observational data
\cite{Alam2004,Alam20041,Alam20042,Alam20043}. Quintessence
\cite{Wetterich1988,Wetterich19881}, phantom (ghost) field
\cite{Cald2002,Cald20021}, K-essence
\cite{Chiba2000,Chiba20001,Chiba20002}, Chaplygin gas
\cite{Kam2001,Kam20011}, holographic dark energy which originates
from quantum gravity
\cite{Li2004,Li20040,Li20041,Sheykhi2010,Sheykhi2011,Sheykhi20102}
and agegraphic DE
\cite{Cai2007,Cai20071,Cai20072,Cai20073,Cai20074,Cai20075,Cai20076,
Cai20077,Sheykhi2009,Sheykhi20091,Sheykhi20092,Sheykhi20093,Sheykhi20094}
are some examples of DE models with time varying EoS parameter. On
the other hand, in another approach, some physicists try to solve
the DE problem by modifying the field equations of GR in such a
way that the phase of acceleration is reproduced without including
any new kind of energy
\cite{Capozziello2003,Capozziello20031,Carroll2004}. Indeed, in
the modified gravity approach, one may consider a new degree(s) of
freedom leading to many unknown features and thus one should
investigate their nature and new consequences in the universe
meaning that this approach adds more complexity to the system.
Therefore, it is impressive and economic if we can explain DE
without entering the new degrees of freedom.

GDE is a model for DE wherein we do not need to introduce new
degrees of freedom or modify gravity. This model is based on the
Veneziano ghost field used in order to solve the so-called U(1)
problem in QCD theory
\cite{Kawa1980,Kawa19801,Kawa19802,Kawa19803,Kawa19804}. Although
there is not any observable consequence from the ghost field in a
Minkowskian spacetime, it produces a small vacuum energy density
proportional to  $\rho_D \sim\Lambda
^3_{QCD}H\sim(3\times10^{-3}eV)^4$, which solves the fine-tuning
problem \cite{Ohta2011}, in curved spacetime. Here,
$\Lambda_{QCD}\sim 100 MeV$ and $H\sim10^{-33}eV$ are QCD mass
scale and Hubble parameter, respectively \cite{Ohta2011}.
Different features of GDE have been studied in ample details
\cite{Ebrahimi2011,Ebrahimi20111,Ebrahimi20112,SheyMov,SheyTav,Chao2012,Chao20121,Chao20122}.
It has been found that the contribution of the Veneziano QCD ghost
field to the vacuum energy is not exactly of order of $H$ and
there is also a second order term proportional to $H^2$ which
contributes to the vacuum energy density \cite{zhit}. Adding the
$H^2$ correction term to the GDE model, one may study the GGDE
model in which the energy density is taken as  $\rho_D=\alpha
H+\beta H^2$ \cite{Cai2012,SheykhiG,SheykhiGBD}.

Based on the cosmological principle, the universe is homogeneous
and isotropic in scales larger than $100$-Mpc and it can be open,
flat or closed denoted by the curvature constant $k=-1,0,1$,
respectively \cite{roos}. It is useful to mention here that
although some observations indicate a flat universe, the nonflat
case is not completely rejected by observations
\cite{roos,Sievers2003,Sievers20031,Sievers20032,Uzan2003,Uzan20031,Uzan20032,Uzan20033,Uzan20034,Uzan20035,Caldwell2005}.
In addition, there are also several observations which indicate a
mutual interaction between DE and dark matter (DM)
\cite{gdo1,gdo2,ob1,ob2,ob3,ob4,ob5,ob6,ob7}. The initial simple
models of the mutual interaction between DE and DM are linear
functions of $\rho_D$ and $\rho_m$
\cite{Amendola1999,Amendola19991,Amendola19992,Amendola19993,Amendola19994,Amendola19995,Amendola19996,Amendola19997,Amendola19998,Amendola19999},
where $\rho_m$ is the energy density of DM.

Moreover, investigations confirm that the sign of the mutual
interaction between DM and DE is changed during the history of
universe \cite{Cai2010}. In this regards, Wei
\cite{WEI2011,Wei2011}. proposed a sign-changeable interaction
term in the form $Q=q(\alpha \dot{\rho}+3\beta H{\rho})$, where
$\alpha$ and $\beta$ are dimensionless constant and $q$ is the
deceleration parameter. It is obvious that the sign of $Q$ is
changed whenever the universe expansion phase is changed from a
deceleration phase $(q>0)$ to an acceleration one $(q<0)$. It is
also worth mentioning that, from the dimensional point of view,
one may consider $\alpha=0$ and discard the $\alpha \dot{\rho}$
term \cite{Wei2011,chimen1,chimen2}. In fact, the sign-changeable
interaction has attracted  a lot of attentions
\cite{Khurshudyan2015,Khurshudyan20151,Khurshudyan20152,Khurshudyan20153,Khurshudyan20154,Khurshudyan20155,Khurshudyan20156,Y.D.
Xu20141,Y.D. Xu2014,Abdollahi2016}. For example, the Chaplygin gas
model of DE with sign-changeable interaction has been investigated
widely in the literatures
\cite{Khurshudyan2015,Khurshudyan20151,Khurshudyan20152,Khurshudyan20153,Khurshudyan20154,Khurshudyan20155,Khurshudyan20156}.
The agegraphic and new agegraphic models of DE with the
sign-changeable interaction have also been explored, respectively,
in \cite{Y.D. Xu20141} and~\cite{Y.D. Xu2014}. Very recently, we
have studied the holographic DE model with the sign-changeable
interaction term with various IR cutoffs \cite{Abdollahi2016}.

In the present paper, we are interested in studying the effects of
considering a mutual sign-changeable interaction between DM and
the DE candidates, including GDE and GGDE, on the evolution
history of universe. Indeed, we are going to investigate how a
sign-changeable interaction affects the description of GDE and
GGDE models of DE about the current phase of the cosmic expansion.
We also investigate the evolution of the system parameters, such
as the equation of state (EoS) parameter as well as the
deceleration and dimensionless density parameters, during the
cosmic evolution from the matter dominated era to the current
accelerating epoch. In order to present our work, we organize the
paper according to the following sections. In section
$\textmd{II}$, we study GDE with the sign-changeable interaction
in both flat and nonflat universes. Thereinafter, we extend our
study to the sign-changeable interacting GGDE in both the flat and
nonflat universes in section $\textmd{III}$ and investigate the
cosmological implications of the model. In section $\textmd{IV}$,
we compare the EoS parameter of the sign-changeable interaction
GDE and the standard GDE model. We summarize our results in
section $\textmd{V}$.

\section{GDE with the sign-changeable interaction}
In this section, we study the GDE  in the presence of the
sign-changeable interaction  term  in both flat and nonflat
universe.
\subsection{Flat Universe}
\begin{figure}[htp]
\begin{center}
\includegraphics[width=8cm]{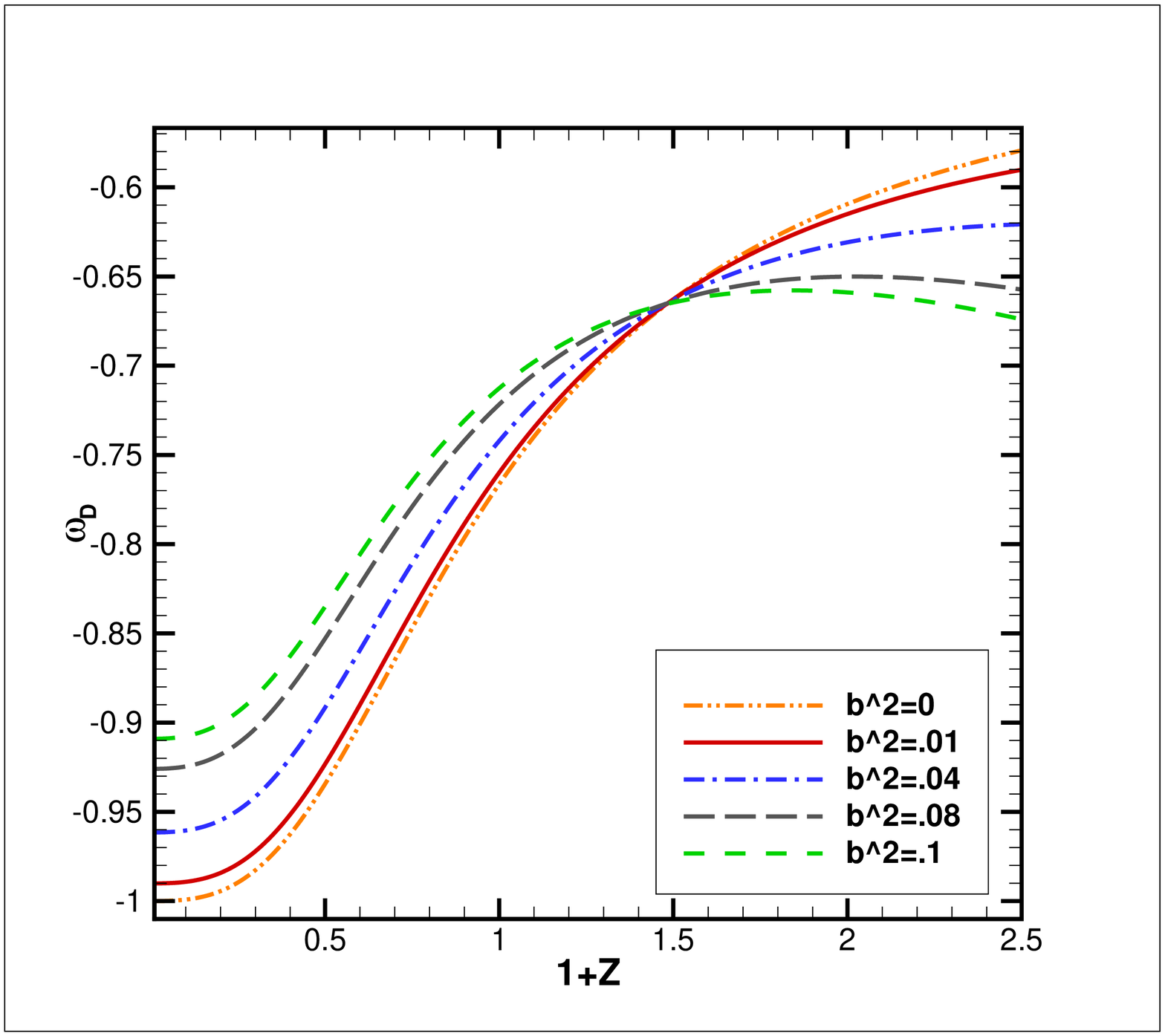}
\caption{The evolution of $\omega_D$ versus redshift parameter $z$
for the sign-changeable interacting GDE in flat
universe.}\label{Eos-z1}
\end{center}
\end{figure}

The first Friedmann equation in a flat homogeneous and isotropic
FRW universe is written as \cite{roos}
\begin{equation}\label{Fri1}
H^2=\frac{8\pi G}{3}(\rho_m+\rho_D),
\end{equation}
where $\rho_D$ is the GDE density and $\rho_m$ is the energy
density of DM. For the GDE density we have \cite{Ohta2011}
\begin{equation}\label{DE2}
\rho_D=\alpha H,
\end{equation}
where $\alpha$ is a constant of order $\Lambda ^3_{QCD} $ and
$\Lambda_{QCD}$ is the QCD mass scale \cite{Ohta2011}. The
fractional energy density parameters and the energy density ratio
are defined as
\begin{equation}\label{Omega}
\Omega_m=\frac{\rho_m}{\rho_{cr}}=\frac{8\pi G\rho_m}{3H^2},\
 \   \  ~~\Omega_D=\frac{\rho_D}{\rho_{cr}}=\frac{8\pi G\alpha
 }{3H},
\end{equation}
and
\begin{equation}\label{r}
r =\frac{\rho_m }{\rho_D
}=\frac{\Omega_m}{\Omega_D}=\frac{1-\Omega_D}{\Omega_D}.
\end{equation}
For an interacting universe in which there is a mutual interaction
between dark sectors of cosmos, the energy-momentum conservation
law can be written as
\begin{eqnarray}\label{con1}
&&\dot{\rho}_m+3H\rho_m=Q,\\
&&\dot{\rho}_D+3H(1+\omega_D)\rho_D=-Q.\label{con2}
\end{eqnarray}
In the above equations, $Q$ denotes the interaction term between
DE and DM. Here, we consider the interaction term as
\cite{Wei2011,Cai2010}
\begin{eqnarray}\label{Q}
Q = 3\beta Hq(\rho_D+\rho_m),
\end{eqnarray}
where $\beta$ is the coupling constant of interaction $Q$, and $q$
is the deceleration parameter defined as
\begin{equation}\label{q}
q=-1-\frac{\dot{H}}{H^2}.
\end{equation}
Let us note that although some negative values are allowed for the
coupling constant $\beta$, we only focus on the $\beta=b^2>0$ case
\cite{Wei2011,Cai2010}. Taking the time derivative of
relation~(\ref{DE2}) and considering Eq.~(\ref{Fri1}), we obtain
\begin{equation}\label{dotDE2}
\dot{\rho}_D=\rho_D \frac{\dot{H}}{H}=-4\pi G\alpha{\rho}_D(1+r+\omega_D).
\end{equation}
Substituting Eqs.~(\ref{dotDE2}) and~(\ref{Q}) into
Eq.~(\ref{con2}) and bearing Eq.~(\ref{r}) in mind, one reaches at
\begin{equation}\label{Eos2}
\omega_D=-\frac{1}{2-\Omega_D}\left(1+\frac {2b^2q}{\Omega_D}\right).
\end{equation}
If we set $q=1$ in Eqs.~(\ref{Q}) and~(\ref{Eos2}), then $Q$ and
$\omega_D$ are reduced to relations obtained in
Ref.~\cite{SheyMov}. In Fig.~\ref{Eos-z1}, considering the initial
condition $\Omega_D(z=0)=0.72$, the evolution of $\omega_D$ is
plotted against the redshift parameter $z$. Intersetingly, the EoS
parameter of the sign-changeable interacting GDE cannot cross the
phantom divide ($\omega_D=-1$) at the late time where
$\Omega_D\rightarrow1$. This is due to the fact that at the late
time $q$ becomes negative and hence $w_D=-(1+2b^2 q)\geq-1$. This
is in contrast to the case of standard interacting GDE, where in
the late time the EoS parameter of interacting GDE necessary
crosses the phantom line, namely, $w_D=-(1+2b^2)<-1$ independent
of the value of coupling constant $b^2$ \cite{SheyMov}. For
example, taking $\Omega_D=0.72$ for the present time, the phantom
crossing take places provided $b^2>0.1$ \cite{SheyMov}.
Using Eqs.~(\ref{q}) and (\ref{dotDE2}), we find
\begin{equation}\label{dotH10}
\frac{\dot{H}}{H^2}=-\frac{3}{2} {\Omega_D}({1+r+\omega_D}),
\end{equation}
which can be combined with Eqs.~(\ref{Eos2}) and~(\ref{q}) to
reach at
\begin{equation}\label{q2}
q=\left(\frac{1}{2}-\frac{3\Omega_D}{2(2-\Omega_D)}\right)\left[\frac{2-\Omega_D}{2-\Omega_D+3b^2}\right].
\end{equation}
\begin{figure}[htp]
\begin{center}
\includegraphics[width=8cm]{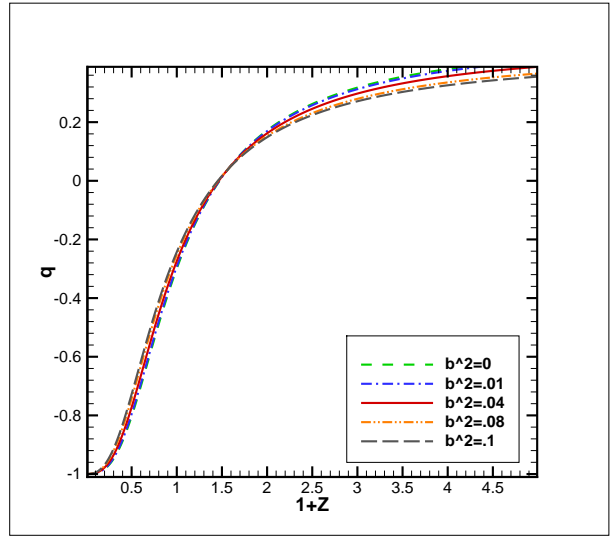}
\caption{The evolution of $q$ versus redshift parameter $z$ for
the sign-changeable interacting GDE in flat universe.}\label{q-z1}
\end{center}
\end{figure}
Considering $\Omega_D(z=0)=0.72$ for the initial condition, we have
plotted $q$ against the redshift parameter in Fig.~\ref{q-z1}. As
it is obvious, there is a transition from the deceleration phase
to the acceleration one at $z\approx 0.6$.
\begin{figure}[htp]
\begin{center}
\includegraphics[width=8cm]{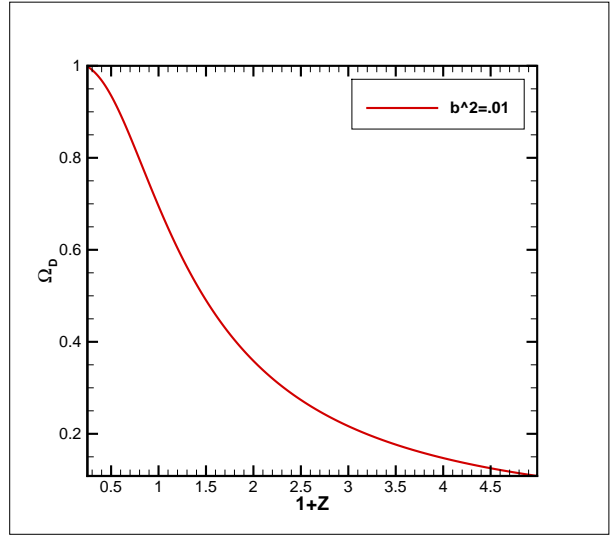}
\caption{The evolution of $\Omega_D$ versus redshift parameter $z$ for
the sign-changeable interacting GDE in flat universe  when $\Omega_D(z=0)=0.72$.}\label{Omega-z1}
\end{center}
\end{figure}
Taking the derivative
with regard to time from $\Omega_D=({8\pi G\rho_D })/{3H^2}$ and
combining the result with Eqs.~(\ref{con2}) and~(\ref{dotH1}), one
can find
\begin{equation}\label{dotOmega1}
\frac{d\Omega_D}{d\ln a}=3\Omega_D\left[\frac{1-\Omega_D}
{2-\Omega_{D}}\left(1+\frac{2b^2q}{\Omega_{D}}\right)-\frac{b^{2}q}{{\Omega}_D}\right]=\frac{3}{2} \Omega_D\left[1-\frac{
\Omega_D}{2- \Omega_D}\left(1+\frac{2b^2q}{\Omega_D}\right)\right].
\end{equation}
We have plotted the dynamics of dimensionless GDE density in
Fig.~\ref{Omega-z1}. We observe that at the early time
$\Omega_D\rightarrow 0$ and at the late time $\Omega_D\rightarrow
1$, as expected. It is easy to check that, as previous, the result
of Ref.~\cite{SheyMov} are obtainable when $q=1$. In summary, for
the sign-changeable interacting GDE in flat universe, at the late
time where $z\rightarrow0$, we have $q\rightarrow-1$ and
$\omega_D\geq-1$.

\subsection{Nonflat Universe}
Here we consider the sign-changeable interacting GDE in a nonflat
universe. It has been argued that the flatness is not a necessary
consequence of inflation if the number of e-folding is not very
large \cite{Huang2004}. The spatial curvature made a contribution
to the energy components of cosmos which is constrained as
$-0.0175<\Omega_k< 0.0085$ with 95\% confidence level by current
observations \cite{Waterhouse}. The first Friedmann equation in a
nonflat homogeneous and isotropic FRW universe is
\begin{equation}\label{Fri2}
H^2+\frac{k}{a^2}=\frac{8\pi G}{3}(\rho_m+\rho_D),
\end{equation}
where $k=-1,0,1$ is the curvature parameter corresponding to open,
flat, and closed universes, respectively. The curvature fractional
density parameter is defined as $\Omega_k={k}/{(a^2H^2)}$, and
thus the Friedmann equation can be rewritten in the following form
\begin{equation}\label{Omegak}
\Omega_m+\Omega_D=1+\Omega_k,
\end{equation}
which also yields
\begin{equation}\label{r1}
r =\frac{\Omega_m}{\Omega_D}=\frac{1+\Omega_k-\Omega_D}{\Omega_D},
\end{equation}
for the energy density ratio. Combining the time derivative of
Eq.~(\ref{Fri2}) with Eq.~(\ref{Omegak}), we obtain
\begin{equation}\label{dotH20}
\frac{\dot{H}}{H^2}=\Omega_k-\frac{3}{2}{\Omega_D} ({1+r+\omega_D}).
\end{equation}
Inserting the above relation into Eq.~(\ref{con2}) and using Eqs.~
(\ref{Q}) and (\ref{dotDE2}), we reach at
\begin{equation}\label{Eos3}
\omega_D=-\frac{1}{2-\Omega_D}\left(1-\frac
{\Omega_k}{3}+\frac{2qb^2}{\Omega_D}(1+\Omega_k)\right),
\end{equation}
\begin{figure}[htp]
\begin{center}
\includegraphics[width=8cm]{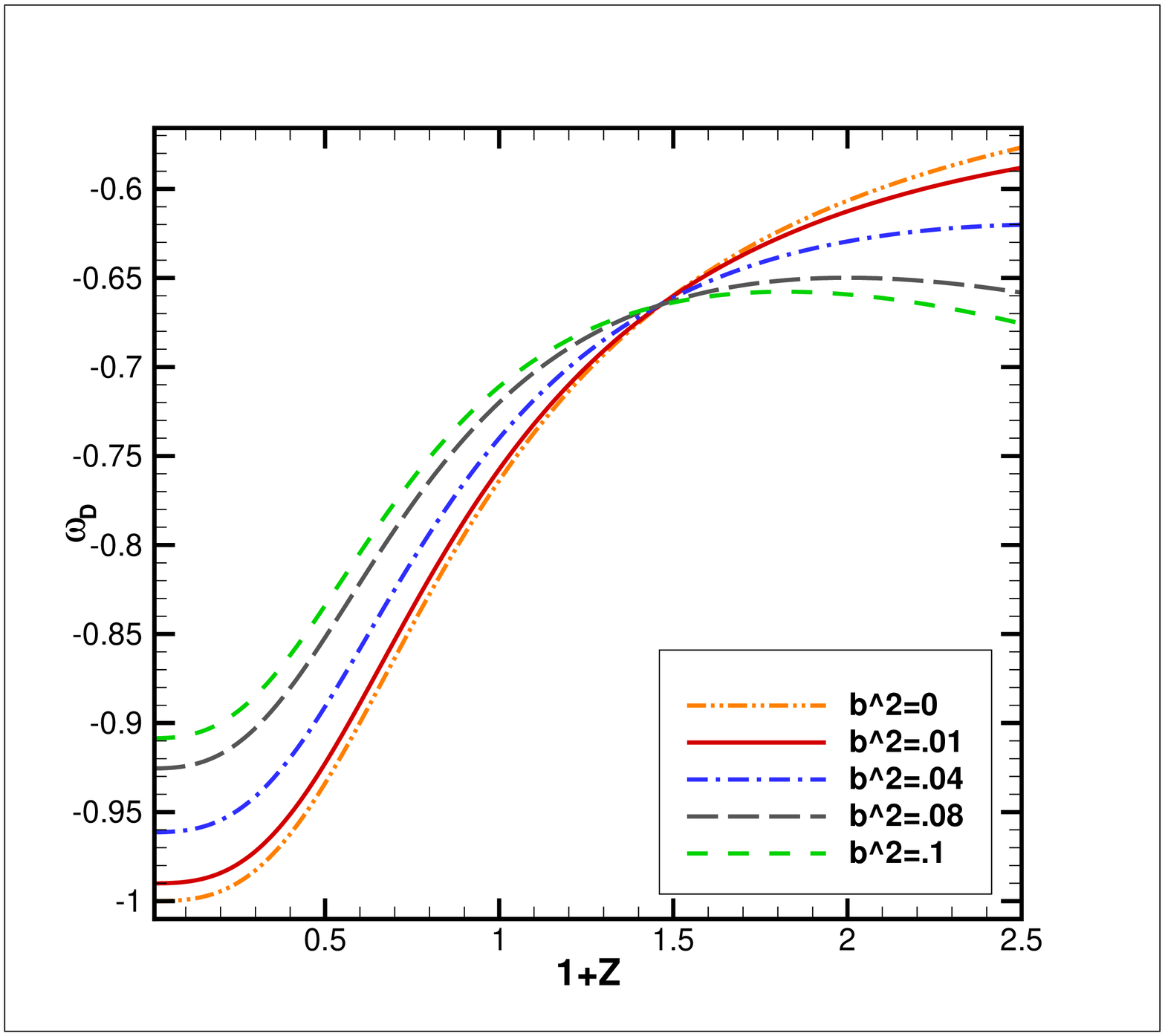}
\caption{The evolution of $\omega_D$ versus redshift parameter $z$
for GDE in a nonflat universe when $\Omega_D(z=0)=0.72$
 and $k=1$.}\label{Eos-z2}
\end{center}
\end{figure}
for the EoS parameter of sign-changeable interacting GDE in a
nonflat universe. Substituting Eq.~(\ref{Eos3}) and~(\ref{dotH20})
into~(\ref{q}), the deceleration parameter in a nonflat background
is obtained as
\begin{equation}\label{q3}
q=\left[\frac{1+\Omega_k}{2}-\frac{3\Omega_D}{2(2-\Omega_D)}\left(1-\frac{\Omega_k}{3}\right)\right]\left(\frac{2-\Omega_D}{2-\Omega_D+3b^2(1+\Omega_k)}\right).
\end{equation}

\begin{figure}[htp]
\begin{center}
\includegraphics[width=8cm]{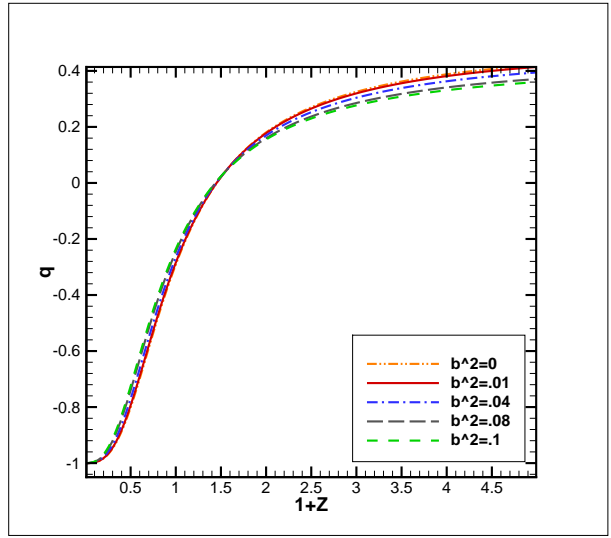}
\caption{The evolution of $q$ versus redshift parameter $z$ for
GDE in a nonflat universe. Here we have taken $\Omega^0_D=0.72$ and
  $k=1$.}\label{q-z2}
\end{center}
\end{figure}
We plot the evolution of $\omega_D$ and $q$ against the redshift
parameter ($z$) for GDE in the closed universe in
Figs~.\ref{Eos-z2} and~\ref{q-z2}, respectively. Again, we see
that the universe has a phase transition from deceleration to an
acceleration around $z\approx 0.6$.
\begin{figure}[htp]
\begin{center}
\includegraphics[width=8cm]{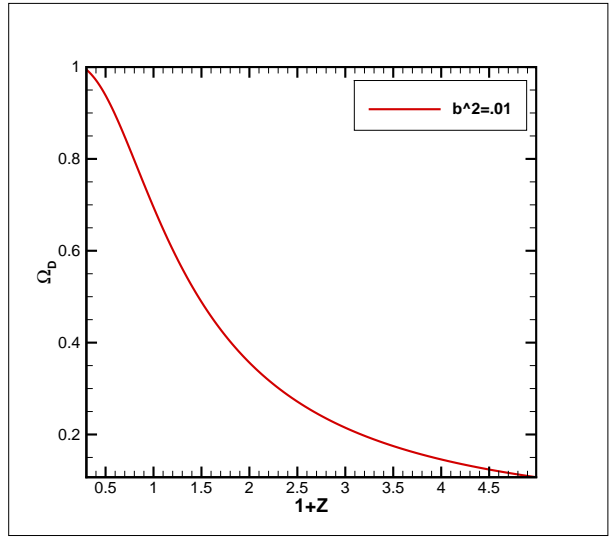}
\caption{The evolution of $\Omega_D$ versus redshift parameter $z$ for
the sign-changeable interacting GDE in nonflat universe when $\Omega_D(z=0)=0.72$ and $k=1$.}\label{Omega-z2}
\end{center}
\end{figure}
 It is a matter of calculation to show that
\begin{equation}\label{dotOmega2}
\frac{d\Omega_D}{d\ln
a}=\frac{3\Omega_D}{2}\left[1+\frac{\Omega_k}{3}-\frac{\Omega_D}{2-\Omega_D}\left(1-\frac{\Omega_k}{3}+\frac{2qb^2}{\Omega_D}(1+\Omega_k)\right)
\right],
\end{equation}
where we used Eqs.~(\ref{dotH20}) and~(\ref{con2}) to get the
above equation. It is worthwhile to mention here that the results
of flat case, obtained in previous subsection, are covered by
setting $\Omega_k=0$. The dynamics of GDE in terms of the redshift
parameter is plotted in Fig.~\ref{Omega-z2}. Clearly, at the early
time it shows $\Omega_D\rightarrow 0$ and at the late time the DE
dominates. In the following we can have $q\rightarrow-1$ and
$\omega_D\geq-1$ at the late time where $z\rightarrow0$.

\section{GGDE with the sign-changeable interaction}
In the previous section, we have assumed the energy density of GDE
as $\rho_D=\alpha H$, while, in general, the vacuum energy of the
Veneziano ghost field in QCD is of the form $H+O(H^2)$
\cite{zhit}. Motivated by the argument given in \cite{mag}, one
may expect that the subleading term $H^2$ in the GDE model might
play a crucial role in the early evolution of the universe, acting
as the early DE. It was shown \cite{Cai2012,SheykhiG,SheykhiGBD}
that taking the second term into account can give better agreement
with observational data compared to the usual GDE. This mode is
usually called the generalized ghost dark energy (GGDE) and our
main task in this section is to investigate the properties of this
model in the presence of the sign-changeable interacting term.
Again, we first consider a flat universe and then generalize our
study to the nonflat case.

\subsection{Flat Universe}
For the energy density of GGDE we have
\begin{equation}\label{DE1}
\rho_D=(\alpha H+\beta H^2),
\end{equation}
where $\beta$ is a constant \cite{zhit,Cai2012}. The fractional
energy density parameters also take the below forms
\begin{equation}\label{Omega}
\Omega_m=\frac{\rho_m}{\rho_{cr}}=\frac{8\pi G\rho_m}{3H^2},\
 \   \  ~~\Omega_D=\frac{\rho_D}{\rho_{cr}}=\frac{8\pi G(\alpha +\beta
 H)}{3H}.
\end{equation}
Here, $\rho_{cr}=\frac{3H^2}{8\pi G}$ denotes again the critical
density. Finally, use~(\ref{Omega}) and~(\ref{DE1}) to obtain
\begin{equation}\label{Omega1}
\frac{4\pi G}{3H}(\alpha+2\beta H)=\frac{\Omega_D}{2}+\frac{4\pi G
\beta}{3}.
\end{equation}
Taking the time derivative of Eq.~(\ref{DE1}), one can find
\begin{equation}\label{dotDE1}
\dot{\rho}_D=\dot{H}(\alpha+2\beta H),
\end{equation}
combined with Eq.~(\ref{Omega}) to reach at
\begin{equation}\label{dotH1}
\frac{\dot{H}}{H^2}=-\frac{3}{2} {\Omega_D}({1+r+\omega_D}),
\end{equation}
finally leading to
\begin{equation}\label{dotH2}
\dot{H}=-4\pi G{\rho_D}({1+r+\omega_D}),
\end{equation}
where $r$ is the energy density ratio~(\ref{r}). Substituting
Eqs.~(\ref{dotDE1}) and~(\ref{Q}) into~(\ref{con2}) and using
Eqs.~(\ref{Omega1}),~(\ref{r}) and~(\ref{dotH2}), we find out
\begin{equation}\label{Eos1}
\omega_D=-\frac{1}{2-\Omega_D-\zeta}\left(1+\frac
{2b^2q}{\Omega_D}-\frac{\zeta}{\Omega_D}\right).
\end{equation}
\begin{figure}[htp]
\begin{center}
\includegraphics[width=8cm]{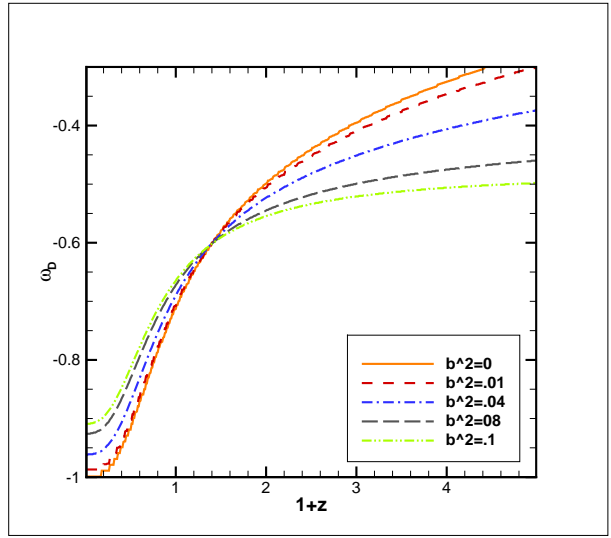}
\caption{The evolution of $\omega_D$ versus redshift parameter $z$ for
the sign-changeable interacting GGDE in flat universe  when $\Omega_D(z=0)=0.72$
, $\zeta=0.1$ .}\label{Eos-z3}
\end{center}
\end{figure}
Here, $\zeta=\frac{8\pi G \beta}{3}$. It is obvious that, as the
flat case, this equation is reduced to the result of
Ref.~\cite{SheykhiG} in the $q=1$ limit. The evolution of
$\omega_D$ has been plotted against the redshift parameter ($z$)
for GGDE in Fig.~(\ref{Eos-z3}).

As the flat case, the EoS of sign-changeable interaction GGDE
cannot cross the phantom division ($\omega_D\geq-1$). Let us note
that at the late time where the universe is  in the accelerated
phase, $q$ becomes negative and considering the fact that
$\zeta=.1$, we arrive at $\omega_D=-\left(1+\frac
{2b^2q}{\Omega_D}-\frac{\zeta}{\Omega_D}\right)\geq-1$. Taking
$q=1$, we have $\omega_D=-\left(1+\frac
{2b^2}{\Omega_D}-\frac{\zeta}{\Omega_D}\right)<-1$, and the result
of Ref.~\cite{SheykhiG} is restored.

Substituting Eq.~(\ref{dotH1}) in~(\ref{q}) and
using~(\ref{Eos1}), one can also obtain
\begin{equation}\label{q1}
q=\frac{1-2\Omega_D+\zeta}{2-\Omega_{D}-\zeta+3b^2}.
\end{equation}
It is easy to verify that the result of Ref.~\cite{SheykhiG} is
covered when $b=0$. Moreover, for $b=0$ and $\zeta=0$, we have
$q=\frac{1-2\Omega_D}{2-\Omega_D}=\frac{1}{2}-\frac{3}{2}{\frac{\Omega_D}{2-\Omega_D}}$
\cite{Ebrahimi20112}.
\begin{figure}[htp]
\begin{center}
\includegraphics[width=8cm]{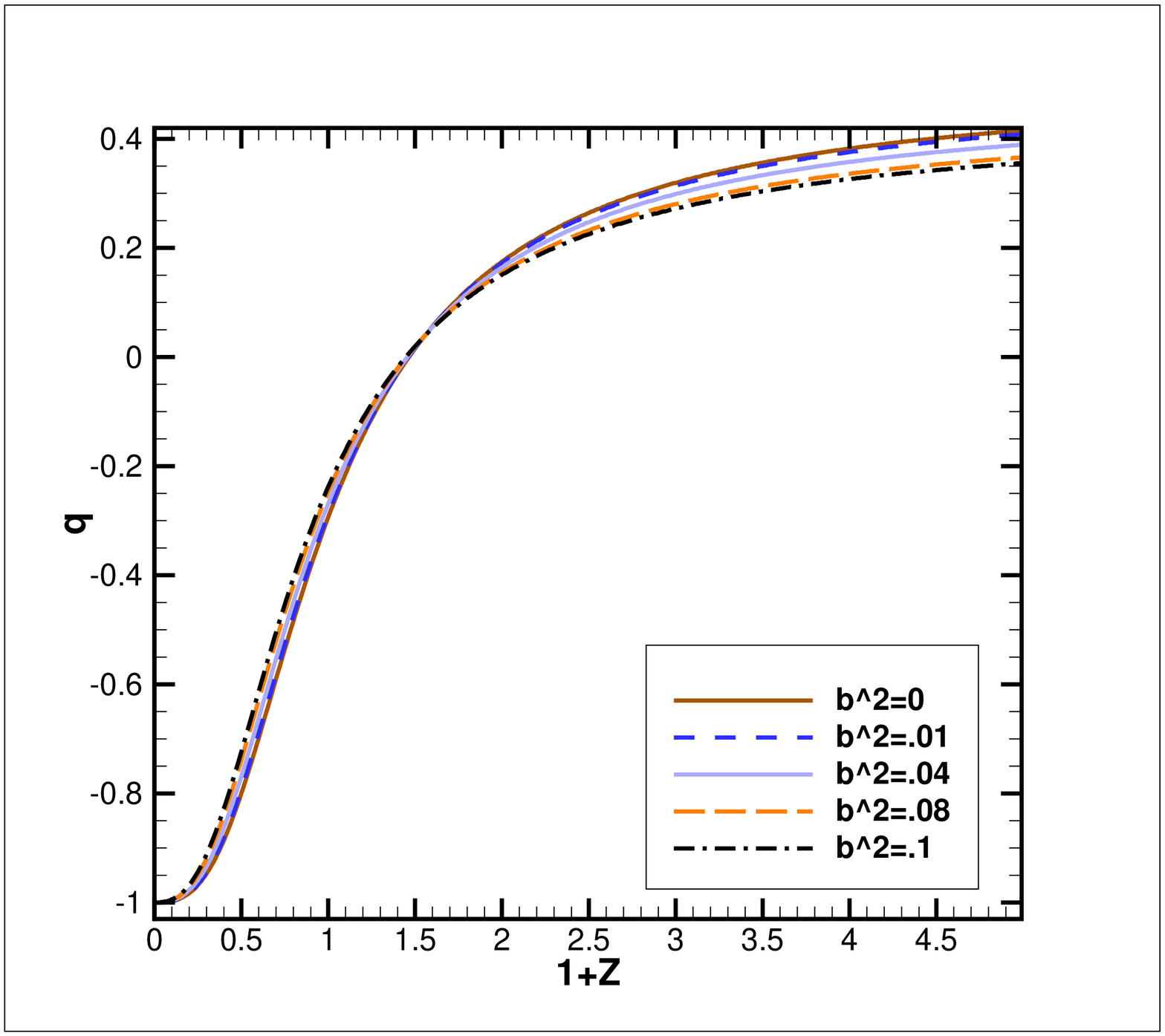}
\caption{The evolution of $q$ versus redshift parameter $z$ for
the sign-changeable interacting GGDE in flat universe  when $\Omega_D(z=0)=0.72$, $\zeta=0.1$ .}\label{q-z3}
\end{center}
\end{figure}
The behavior of $q$ has also been plotted in Fig.~\ref{q-z3},
addressing a transition from the deceleration phase to the
acceleration one at $z\approx 0.6$. Finally, taking the time
derivative of relation $\Omega_D=\frac{8\pi G\rho_D }{3H^2}$ and
using~(\ref{con2})
 and~(\ref{dotH1}), we find
\begin{equation}\label{dotOmega}
{\Omega}^{\prime}_{D}=3\Omega_D\left[\frac{1-\Omega_D}
{2-\Omega_{D}-\zeta}\left(1+\frac{2b^2q}{\Omega_{D}}-\frac{\zeta}{\Omega_D}\right)-\frac{b^{2}q}{{\Omega}_D}\right].
\end{equation}
It is also easy to check that the results of
Refs.~\cite{SheykhiG,Ebrahimi20112} are obtainable from the
above relations.

\begin{figure}[htp]
\begin{center}
\includegraphics[width=8cm]{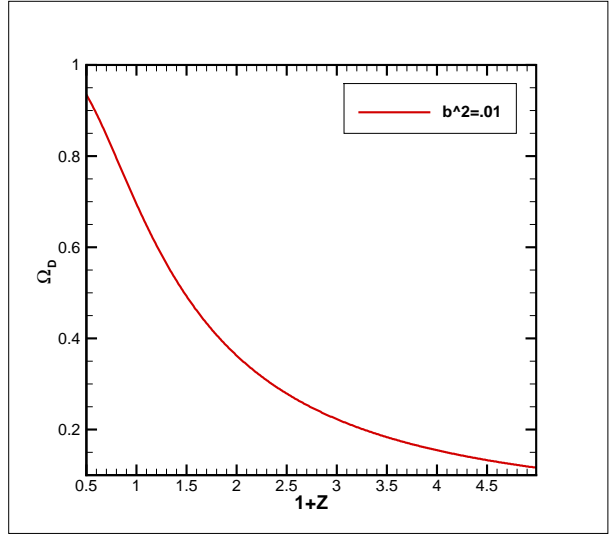}
\caption{The evolution of $\Omega_D$ versus redshift parameter $z$ for
the sign-changeable interacting GGDE in flat universe  when $\Omega_D(z=0)=0.72$
, $\zeta=.1$ .}\label{Omega-z3}
\end{center}
\end{figure}
We have plotted the dynamics of density parameter in
Fig.~\ref{Omega-z3}, and the behavior is similar to the previous
case; at the early time $\Omega_D\rightarrow 0$, while at the late
time $\Omega_D\rightarrow 1$.
\subsection{Nonflat Universe}
In order to find the EoS parameter of sign-changeable interacting
GGDE in the non-flat universe, inserting Eq.~(\ref{dotH2}) into
Eq.~(\ref{dotDE1}) and combining the result with Eqs.~(\ref{con2})
and~(\ref{r1}), we get
\begin{equation}\label{Eos4}
\omega_D=-\frac{1}{2-\Omega_D-\zeta}\left[2-\left(1+\frac
{\zeta}{\Omega_D}\right)\left(1+\frac{\Omega_k}{3}\right)+\frac{2b^{2}q}{\Omega_D}\left(1+\Omega_k\right)\right].
\end{equation}
As one can see the EoS parameter cannot cross the phantom divide
at the late time, because at this epoch we have
$\Omega_D\rightarrow 1$ and $q$ becomes negative, therefore
$\omega_D=-\left(2-(1+\zeta)(1+\frac{\Omega_k}{3})+{2b^{2}q}(1+\Omega_k)\right)
\geq-1$ (note that we have chosen $\zeta = .1$ and
$\Omega_k=.01$). If we set $q=1$ we get
$\omega_D=-\left(2-(1+\zeta)(1+\frac{\Omega_k}{3})+{2b^{2}}(1+\Omega_k)\right)
<-1$, which is the result of Ref.~\cite{SheykhiG}. Thus in
contrast to the EoS parameter of the usual interacting GGDE which
the phantom regime can be achieved, in case of sign-changeable
interaction term the EoS parameter of GGDE is always
$\omega_D\geq-1$.

Combining Eq. (\ref{Eos4}) with Eqs.~(\ref{dotH20}) and~(\ref{q}),
one arrives at
\begin{equation}\label{q4}
q=\left(\frac{1+\Omega_k}{2}-\frac{3\Omega_D}{2(2-\Omega_D-\zeta)}(1-\frac{\Omega_k}{3})\right)\left[\frac{2-\Omega_D}{2-\Omega_D+3b^2(1+\Omega_k)}\right],
\end{equation}
for the deceleration parameter. One can finally use
Eqs.~(\ref{Omega}),~(\ref{con2}) and~(\ref{dotH20}) in order to
obtain
\begin{equation}\label{dotOmega3}
{\Omega}^{\prime}_{D}=3\Omega_D\left[\frac{\Omega_k}{3}+
\frac{1-\Omega_D}{2-\Omega_D-\zeta}\left(2-(1+\frac{\zeta}{\Omega_D})(1+\frac{\Omega_k}{3})+
\frac{2b^{2}q}{\Omega_D}(1+\Omega_k)\right)-\frac{b^{2}q}{\Omega_D}(1+\Omega_k)\right].
\end{equation}
\begin{figure}[htp]
\begin{center}
\includegraphics[width=8cm]{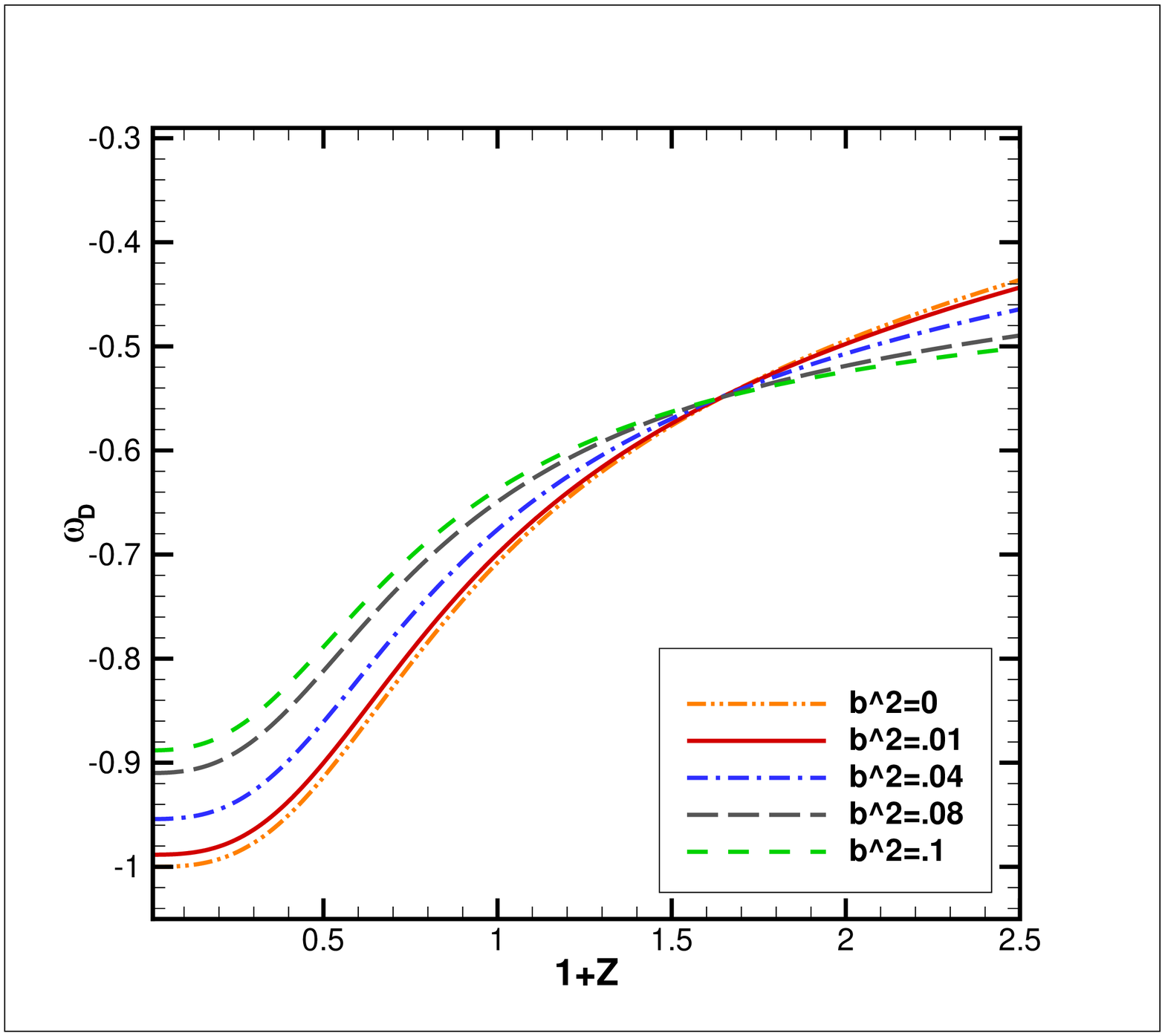}
\caption{The evolution of $\omega_D$ versus redshift parameter $z$ for
the sign-changeable interacting GGDE in nonflat universe when $\Omega_D(z=0)=0.72$,
$\zeta=0.1$ and $k=1$
.}\label{Eos-z4}
\end{center}
\end{figure}

\begin{figure}[htp]
\begin{center}
\includegraphics[width=8cm]{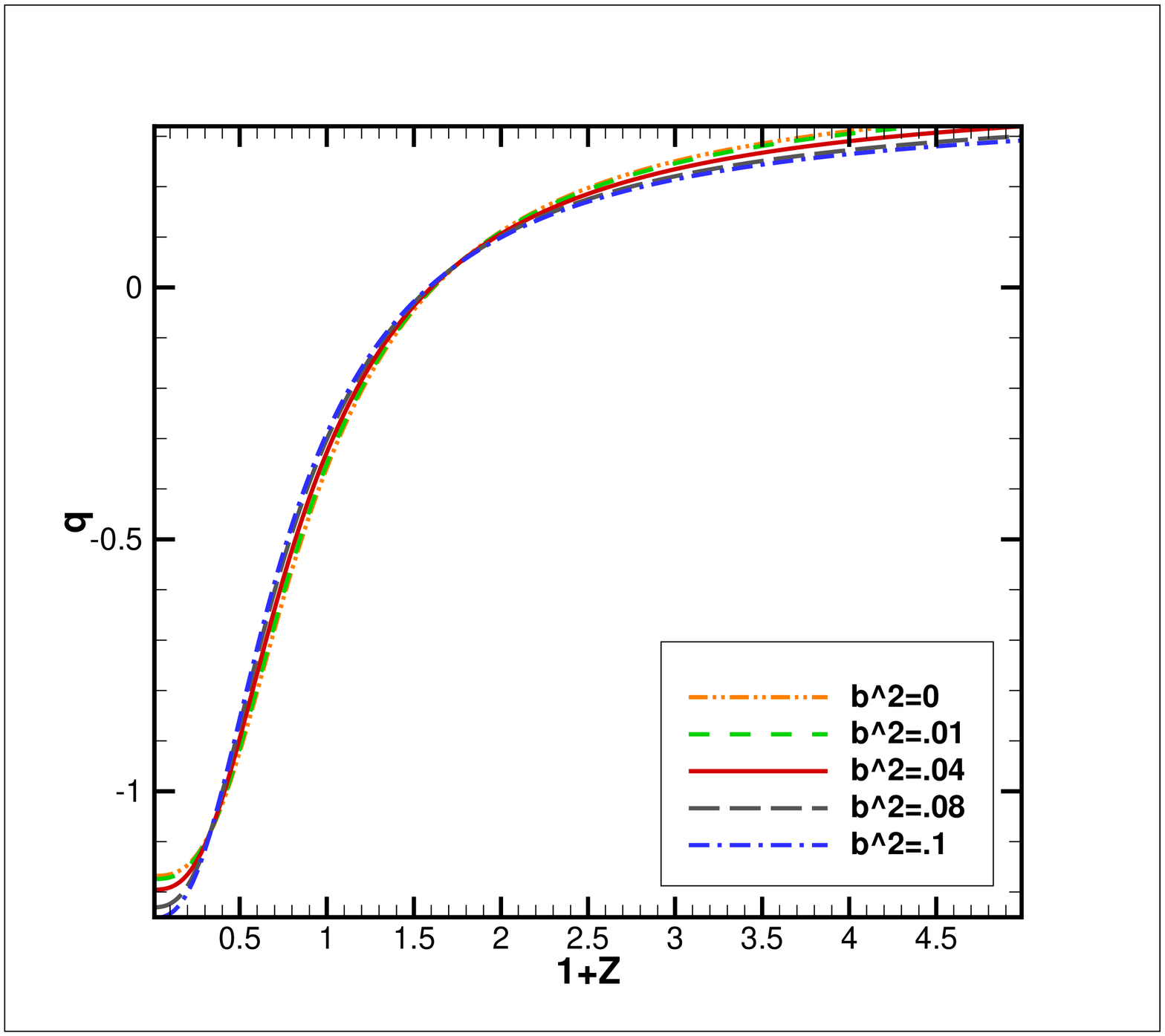}
\caption{The evolution of $q$ versus redshift parameter $z$ for
the sign-changeable interacting GGDE in nonflat universe when $\Omega_D(z=0)=0.72$,
$\zeta=0.1$ and $k=1$
.}\label{q-z4}
\end{center}
\end{figure}

\begin{figure}[htp]
\begin{center}
\includegraphics[width=8cm]{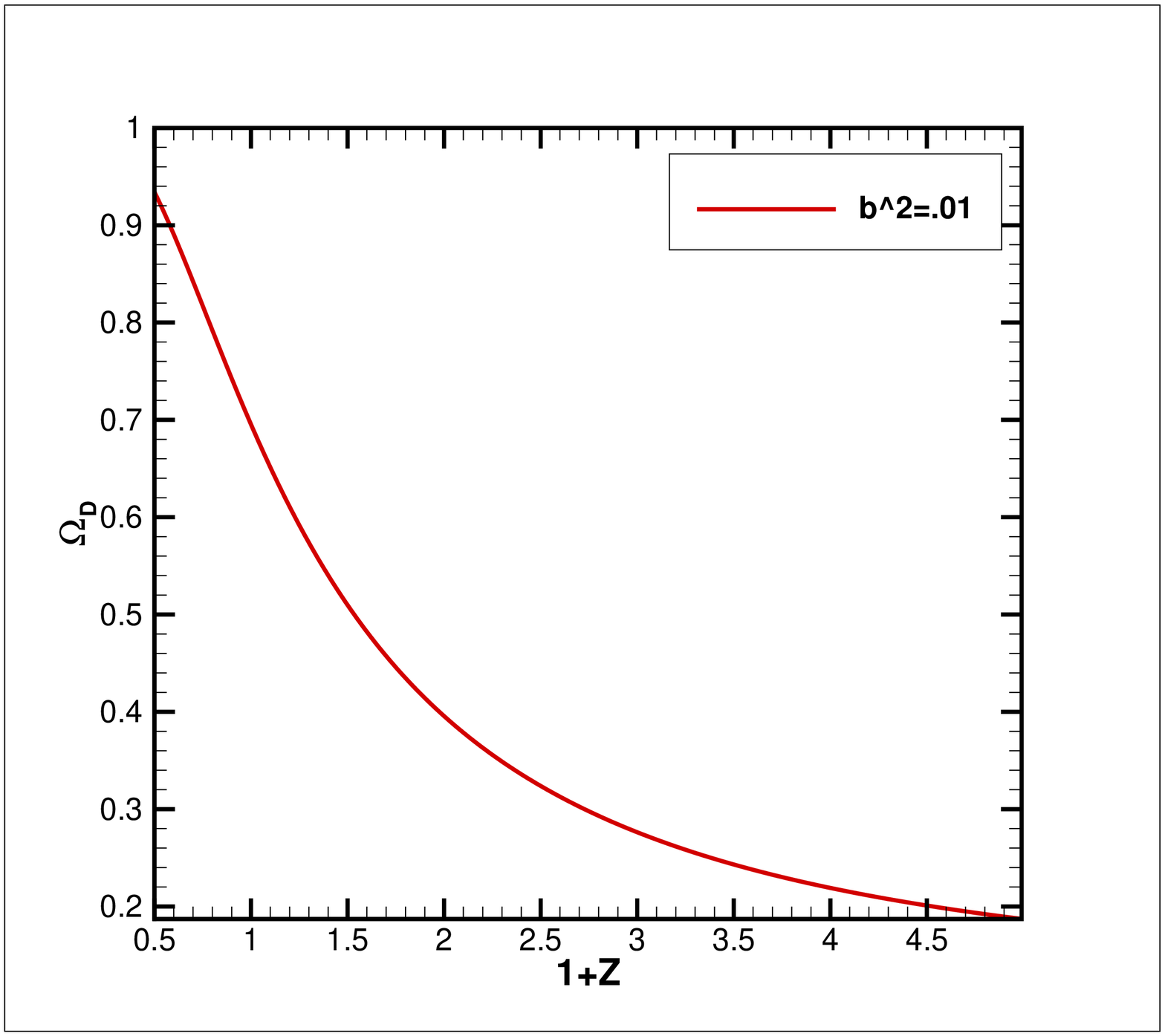}
\caption{The evolution of $\Omega_D$ versus redshift parameter $z$  for
the sign-changeable interacting GGDE in nonflat universe when $\Omega_D(z=0)=0.72$,
$\zeta=0.1$ and $k=1$
 .}\label{Omega-z4}
\end{center}
\end{figure}
It is worth mentioning that in the limit of $\Omega_k=0$, all the
obtained relations in this subsection restore their respective
expressions in the previous subsections  for flat universe. The
behaviors of $\omega_D$ and $q$ against the redshift parameter for
GGDE in the closed universe have also been plotted  in
Figs.~\ref{Eos-z4} and~\ref{q-z4}. The main results of this
figures are: $(i)$ at late time, we have $\omega_D\geq-1$ and
$q<-1$. $(ii)$ there is a transition from the deceleration phase
to the accelerated one around $z\simeq0.6$. We have also plotted
the evolutionary of the GGDE  density in Fig.~\ref{Omega-z4}.

\section{Comparison of EoS parameter of usual interacting GDE and sign-changeable model}
Finally, we compare the original interating GDE model with the
sign-changeable interacting GDE model. For this purpose, we plot
the evolution of $\omega_D$ versus redshift parameter $z$ in Figs.
13 and 14 for both of models GDE and GGDE in a flat and nonflat
universe. The long-dash and dash-dot lines show the evolution of
$\omega_D$ for the sign-changeable interacting GDE model and the
solid and dashed lines show the usual interacting GDE model with
interaction term $Q = 3b^2H(\rho_D+\rho_m)$.  From these figures,
we observe that the EoS parameter of both GDE and GGDE  with
sign-changeable interaction term cannot cross the phantom divide
$\omega_D=-1$ and we always have $\omega_D\geq-1$ at the late
time. In contrast, the EoS parameter of the usual interacting GDE
and GGDE can cross the phantom line, namely $\omega_D<-1$ at the
late time.

\begin{figure}[htp]
\begin{center}
\includegraphics[width=8cm]{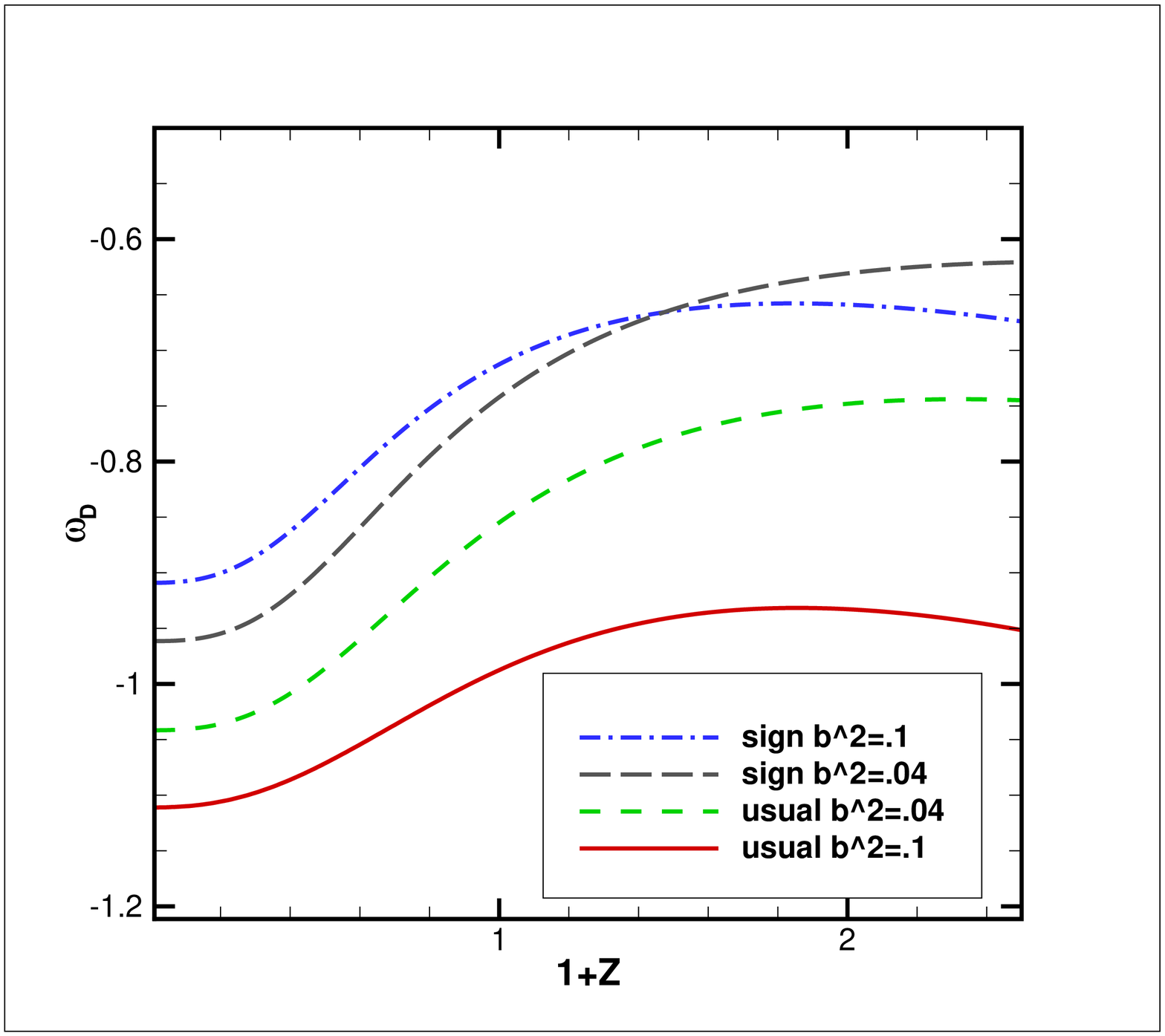}
\includegraphics[width=8cm]{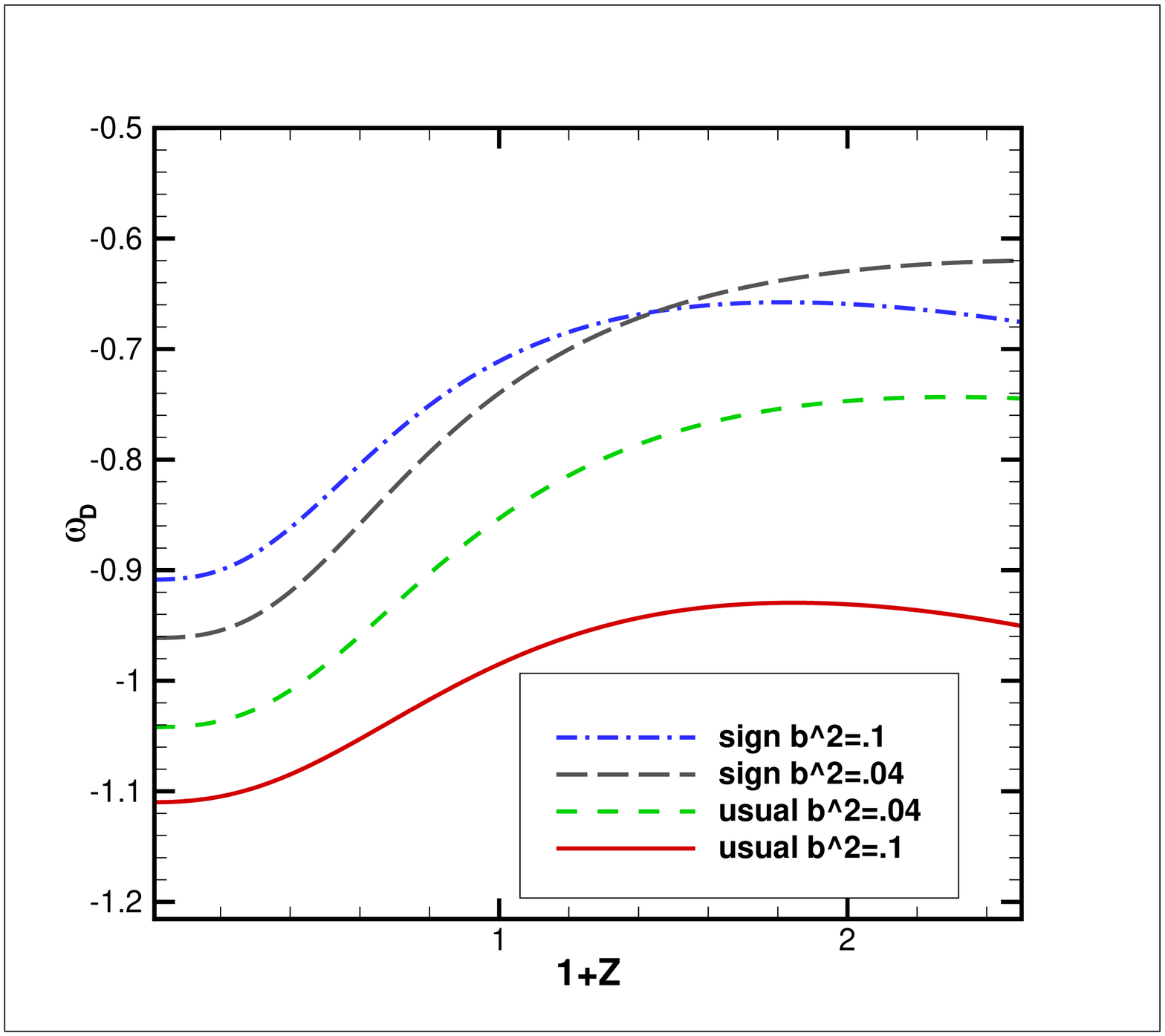}
\caption{The evolution of $\omega_D$ versus redshift parameter $z$
for GDE in a flat and nonflat universe when $b^2=.1,.04$,$\Omega_D(z=0)=0.72$
 and $k=1$.}\label{GDE}
\end{center}
\end{figure}

\begin{figure}[htp]
\begin{center}
\includegraphics[width=8cm]{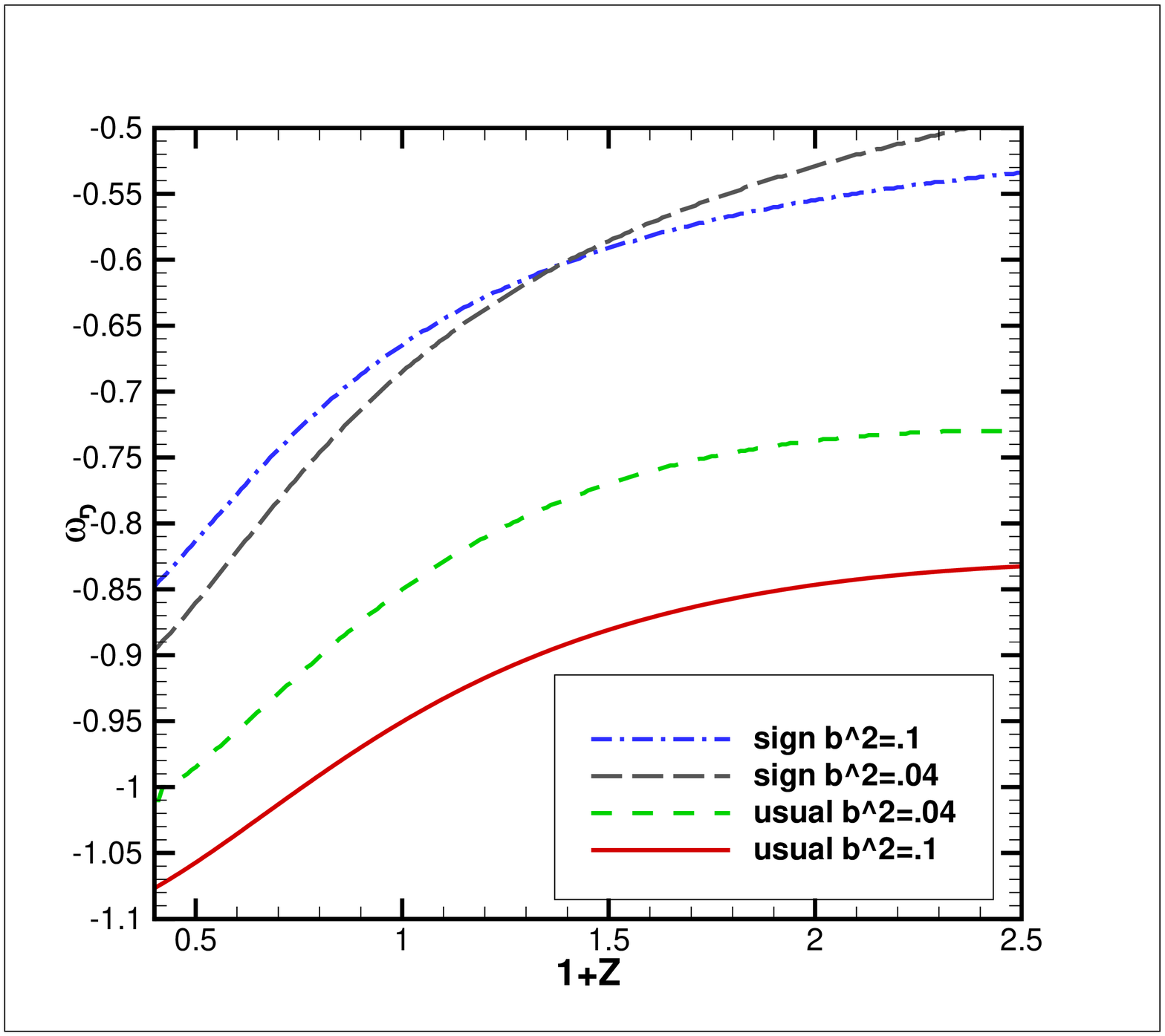}
\includegraphics[width=8cm]{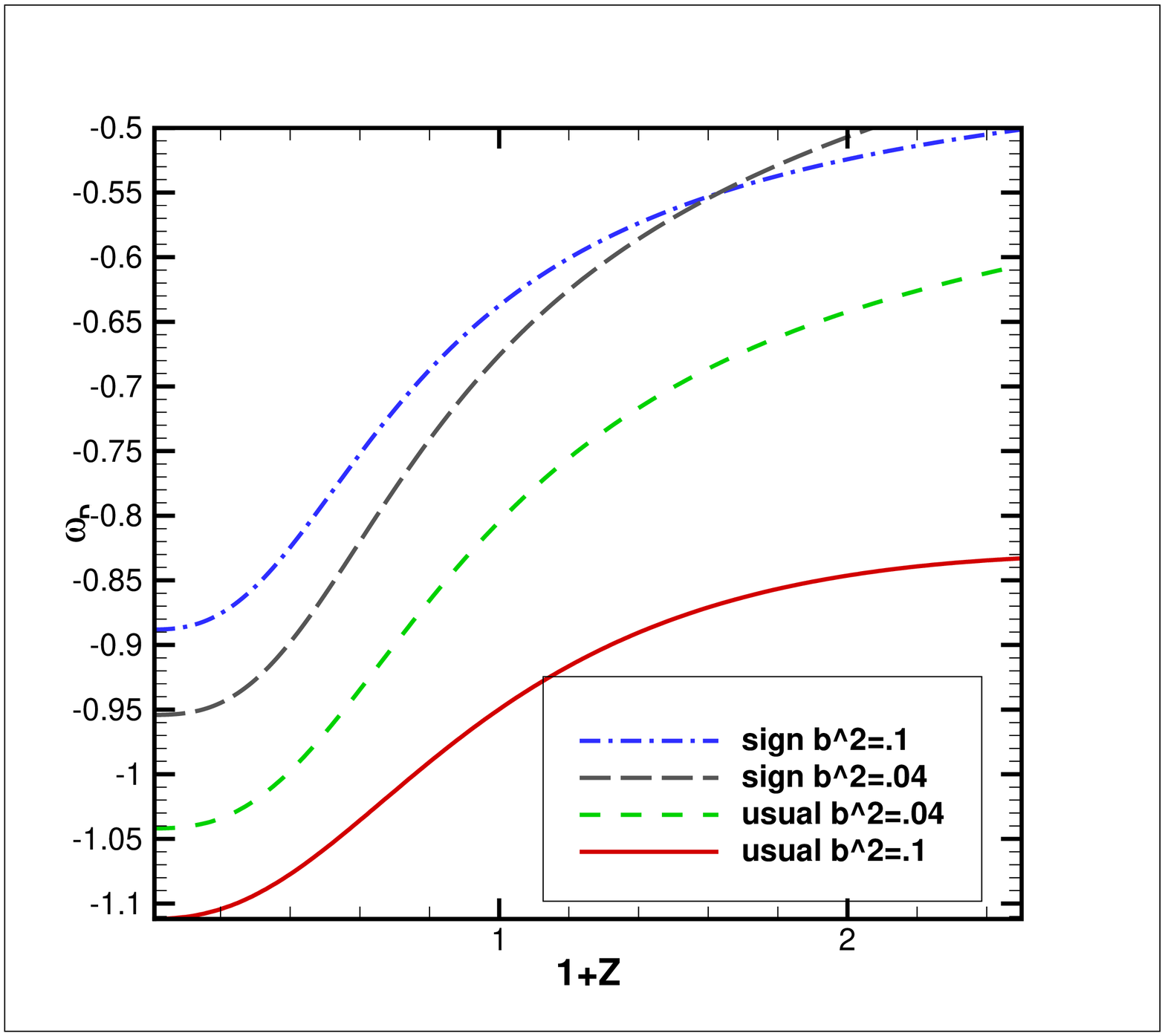}
\caption{The evolution of $\omega_D$ versus redshift parameter $z$
for GGDE in a flat and nonflat universe when $b^2=.1,.04$,$\Omega_D(z=0)=0.72$,$\zeta=0.1$ and $k=1$.}\label{GGDE}
\end{center}
\end{figure}
\section{Closing remarks}
The DE puzzle is undoubtedly one of the most important challenges
of modern cosmology \cite{EDMUND2006}. In this paper, we
considered a flat FRW universe filled by a DM and GDE interacting
with each other through a sign-changeable interaction term. The
generalization to the nonflat case is also investigated, which
shows that, for a closed universe, although $\omega_D\geq-1$ at
late time, we have $q<-1$ for the deceleration parameter. Our
studies show that, at the late time, we have $q\rightarrow-1$ and
$\omega_D\geq-1$ meaning that this model does not cross the
phantom line, a result which is consistent with the cosmological
constant model of DE.

The values of the model parameters can be estimated by fitting the
model with observational data. The observational data for
coefficient $\beta$ in original interaction model, $Q = 3\beta
H(\rho_D+\rho_m)$, implies a positive value ($\beta>0$), hence we
consider $\beta$ to be positive and can be rewritten
$\beta=b^2>0$. We found out that if we select sign-changeable
interaction model, $Q = 3b^2q H(\rho_D+\rho_m)$, because $q$ at
the late time should have a negative value, we cannot have
crossing phantom. Our studies here show that with the
sign-changeable interaction term, only if coefficient $\beta$ in
$Q$ is chosen as a negative value, we can reach the phantom
regime.  All of the studied cases indicate a transition from the
deceleration phase to an accelerated one which take places around
$z\approx 0.6$.
\acknowledgments{We thank Shiraz University Research Council. This
work has been supported financially by Research Institute for
Astronomy \& Astrophysics of Maragha (RIAAM), Iran.}

\end{document}